\documentclass[aps,prl,twocolumn,groupedaddress]{revtex4-2}

\usepackage{amsfonts}
\usepackage{amsmath}
\usepackage{graphicx}
\usepackage[T1]{fontenc}
\usepackage{float}

\begin{document}


\title{A link between shape dependent lifetimes and thermal escape in quantum dots and rings}


\author{H.~T.~Sullivan and J.~H.~Cole}
\affiliation{Chemical and Quantum Physics, and ARC Centre of Excellence in Exciton Science, School of Science, RMIT University, Melbourne 3000, Australia}


\date{\today}

\begin{abstract}
Understanding the optical emission characteristics of semiconductor nanostructures is important when determing their device applicability. The emission depends on the material and its geometry, but also depends on other processes such as thermal escape from the nanostructure. Although it is widely accepted that scattering involving longitudinal optical phonons is the key process in thermal escape, it remains unclear why some quantum structures thermally emit excitons and other single charge carriers. To investigate this phenomena we theoretically determine the energy levels and temperature-lifetime relationships of quantum dots and rings. We find that replicating the observed temperature dependence of the exciton lifetime requires both an eigenspectrum and a thermal escape mechanism which are geometry dependent. This suggests that geometry may be a significant factor in determining the dominant thermal escape process in quantum structures.
\end{abstract}


\maketitle

\section{Introduction}
There is a rapidly growing field of technologies that depend on semiconductor nanostructures \cite{harrison2016quantum, orieux2017semiconductor, michler2017quantum}. Quantum dots (QDs) have been central to these technological developments. The applications for QDs range from biological sensing to single photon sources to quantum computation \cite{medintz2005quantum, senellart2017high, loss1998quantum}. Improvements in fabrication techniques has resulted in the realisation of structures that are more complicated than QDs. These include rods, dumbbells and platelets, to name a few \cite{lan2015nanorod, pisanello2010room, moreels2012nearly, di2015single, ji2017heavy, jia2019heavy, ithurria2011colloidal, guzelturk2014stacking, guzelturk2014amplified}. Each quantum structure has its own unique properties and potential device applications. 

Of particular interest in the field of semiconductor nanostructures is quantum rings (QRs). Whilst exhibiting a range of novel properties, the most significant quantum mechanical effect studied in QRs is the Aharonov-Bohm (AB) effect \cite{aharonov1959significance}. The characteristic AB-oscillations have now been realised in a number of QR systems \cite{lorke2000spectroscopy, bayer2003optical, keyser2002aharonov}. 

The unique properties that QRs offer make them suitable to a number of applications. For instance, broad-area lasers have been constructed out of stacked layers of QRs \cite{suarez2004laser}. A simple scheme involving the use of electric and magnetic fields has been proposed to trap excitons in QRs \cite{fischer2009exciton}. Photonic crystal lattice containing QRs have been demonstrated to act as single photon emitters \cite{gallardo2010single}. QRs have also been of interest to the spintronics community due to their long spin relaxation time \cite{zipper2011spin}. It has been suggested that the additional degrees of freedom in QRs, arising from their geometry, could be utilised for qubit manipulation \cite{zipper2006flux}.

An important question pertaining to QRs is explaining their peculiar relationship between temperature and exciton lifetime. When the temperature is increased above 150K, excitons in QRs exhibit a sudden increase in lifetime \cite{lin2009temperature, lin2009shape}. This contrasts with the typically observed behaviour of QDs which undergoes a shortening of excitonic lifetime as the temperature is increased \cite{heitz1999temperature, de2001capture, valerini2005temperature, de2006size}. This phenomena has been explained by assuming the many dark states that are energetically close to the bright ground state become thermally populated at higher temperatures. A similar temperature-lifetime relationships has been observed in quantum wells and was also attributed to dark states \cite{badcock2013evidence}.

\begin{figure}[ht!]
	\includegraphics[width=\linewidth]{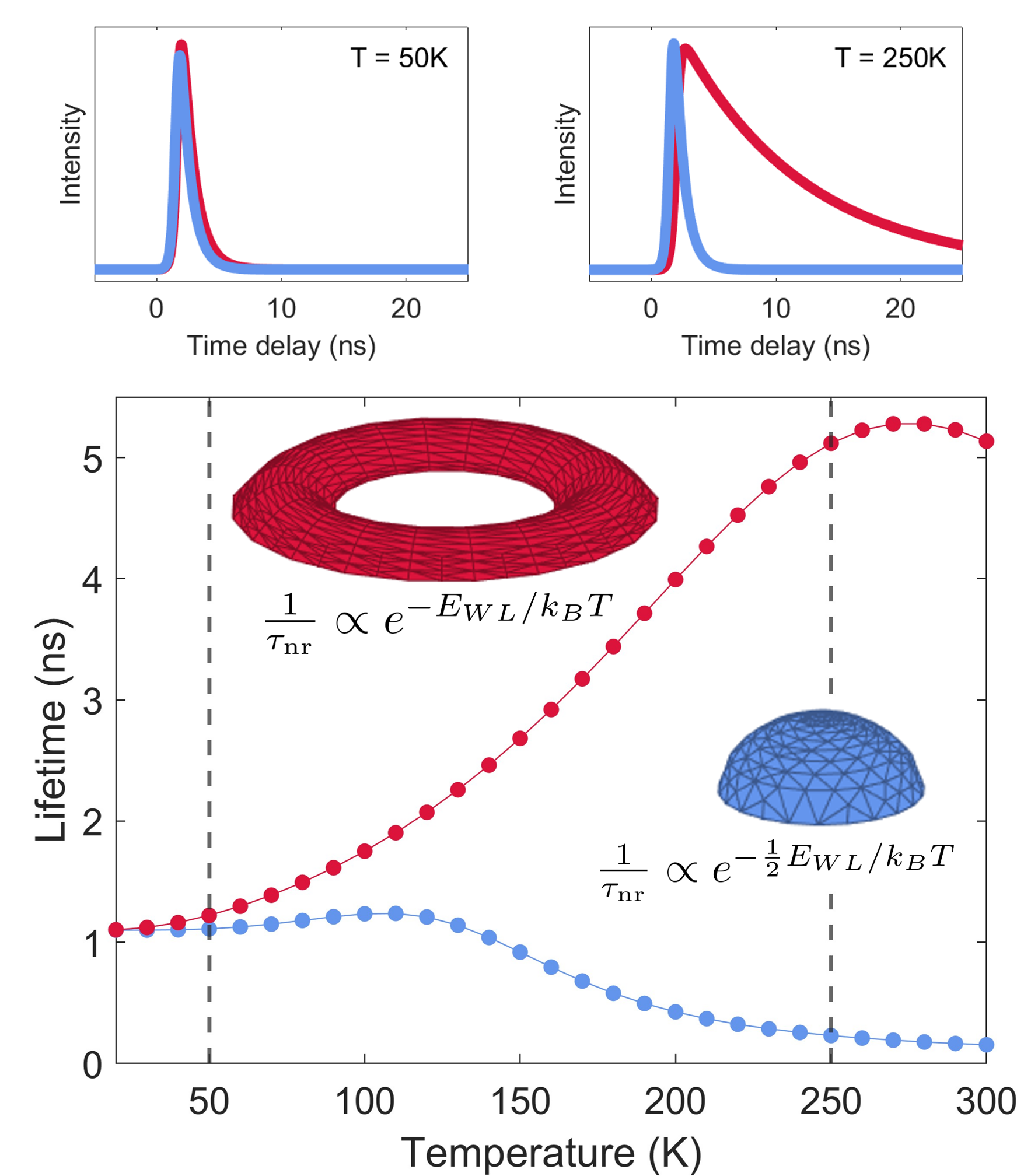}
	\caption{Conceptual diagram illustrating the temperature dependence of exciton lifetime for quantum dots and quantum rings. The temperature-lifetime relationships of QRs (red) and QDs (blue) are significantly different. The QR undergoes a significant rise in lifetime as temperature is increased whereas the QD experiences a declining lifetime. Examples of the time-resolved photoluminesce spectroscopy typically seen at 50K and 250K are demonstrated above the main figure. This is the measurement that determines the lifetime of the QR and QD. The non-radiative rates ($\frac{1}{\tau_{\text{nr}}}$) scales as $\exp(-\nu E_{\text{WL}}/ k_B T)$, where $\nu$ is a parameter that can be determined from the temperature-lifetime relationship. To match published experiments we find that for QRs $\nu = 1$ which corresponds to exciton emission as the dominant thermal escape mechanism. For the QDs we find $\nu= 1/2$ which implies thermal emission via a correlated eletron-hole pairs.}
	\label{fig:fig1}
\end{figure}

Alternatively the temperature-lifetime relationship observed in QRs could be due to the piezoelectric effect \cite{fomin2013physics}. It is argued that the strain in QRs could induce a significant piezoelectric potential which reduces the overlap between the electron and hole wavefunction. For temperatures above 150K these weakly overlapping states may become occupied and, therefore, an increasing lifetime would be observed.

Neither of these explanations take into account thermal escape from the QR and how this would effect its temperature-lifetime relationship. Whether thermal escape is mediated through the ejection of excitons, correlated electron-hole pairs or as single charge carriers has been shown to have a significant effect on the temperature-lifetime relationship in QDs \cite{yang1997effect}. Whilst it is not well understood, it is relatively simple to determine which escape mechanism is dominant in any particular systems and it has been resolved for a variety of QDs \cite{yang1997effect, khatsevich2005cathodoluminescence, schulz2009optical, gelinas2010carrier, jahan2013temperature}. To our knowledge this property has not been determined for QRs. 

In this paper, we employ the finite element method (FEM) to model QDs and QRs that were fabricated by Lin \textit{et al.} \cite{lin2009temperature, lin2009shape}. The FEM allows for the specific geometry of each structure to be included in our calculations permitting a meticulous assessment of the influence that the shape of quantum structures has on their energies and wavefunctions. Performing calculations that are based on the actual geometry of the QRs, and not an idealised shape, allows us to determine the selection rules of this sytem and explain the observed PL spectra.

The states we calculate are used to determine the temperature-lifetime relationship of the QDs and QRs. We demonstrate that this relationship is highly sensitive to the energy and spacing of the states in each structure and, therefore, is senstive to the geometry of the system. We also find a shape dependence in the type of thermal emission from each structure. Thus geometry, through a quantum structure's spectrum of states and thermal emission mechanism, is found to have a significant effect on the lifetime of quantum systems.

\section{Modelling quantum dots and quantum rings} \label{sec:FEM}
In this section we describe how we model QDs and QRs using the finite element method (FEM). The strength of this technique is that it allows differential equations to be solved across arbitrary geometries. The flexibilty that FEM offers in modelling different shapes has resulted in it being used to investigate a range of quantum structures \cite{melnik2004finite, markussen2006influence, fomin2007theory}.

The set of differential equations we use the FEM to solve is the Schr\"{o}dinger-Poisson (SP) equations. These coupled equations are solved iteratively to account for Coulomb interaction between the electron and hole in excitons. The method involves first solving the Schr\"{o}dinger equation to determine the wavefunction of each particle, from which the charge density can be computed. This becomes an input into the Poisson equation which is solved to determine the Coulombic potential that each particle generates. These potentials are then fed back into the Schr\"{o}dinger equation. The two equations are repeatedly solved until the energies converge, 
\begin{subequations}
\begin{align}
	\bigg( -\frac{\hbar^2}{2 m_e} \nabla^2 + V_e(\mathbf{r})  - q |\phi_h| \bigg)\psi_e(\mathbf{r})  &= E_e\psi_e(\mathbf{r}) \\
	\bigg( -\frac{\hbar^2}{2 m_h} \nabla^2 + V_h(\mathbf{r})   - q| \phi_e| \bigg)\psi_h(\mathbf{r})  &= E_h\psi_h(\mathbf{r})  
\end{align}
\end{subequations}
\begin{subequations}
	\begin{align}
	\nabla^2 \phi_{e}   = -|q| \frac{ |\langle \psi_{e} | \psi_{e} \rangle|^2 }{\varepsilon_0 \varepsilon_r} \\
	\nabla^2 \phi_{h}   = -|q| \frac{ |\langle \psi_{h} | \psi_{h} \rangle|^2 }{\varepsilon_0 \varepsilon_r}
\end{align}
\end{subequations}
where the effective mass, wavefunction, energy and Coulomb potential of the electron (hole) is $m_{e(h)}$, $\psi_{e(h)}$, $E_{e(h)}$ and $\phi_{e(h)}$. The potential that the electron (hole) resides in $V_{e(h)}$ arises from the heterojunction between the quantum structure and the surrounding material. The magnitude of charge of the electron and hole is $q$. The vacuum and relative permittivity are $\varepsilon_0$ and $\varepsilon_r$, respectively.

Employing the FEM requires defining a geometry for each structure. We commence by modelling the QD as it has a particuarly simple structure. The QDs that Lin experiments on are estimated to have an average radius and height of 10nm and 2nm, respectively, and they measure the ground state exciton energy to be 1210meV \cite{lin2009shape}. We simulate the QD by assuming it is approximately lens shaped with the following height profile $h(r) = h_0\cos(\pi r / 2 R)^{1/2}$ where $R$ is radius of the QD and $h_0$ is its maximum height. We find that QDs with the same radius as given by Lin but a maximum height $h_0 = 2.35$nm results in a ground state energy of 1210eV. The greater height required in our simulations may be attributed to the uncertainty in the QDs dimensions or the extra volume provided by the wetting layer.  


Modelling QRs is difficult due to the unevenness in their geometry. This unevenness is largely due to their complicated growth process. QRs are produced by the partial capping of a QD followed by an annealing process that triggers the dot-to-ring transition. This transition has been attributed to the balance of surface free energies on the structure \cite{blossey2002wetting, kobayashi2004self}. During this transformation, a redistribution of material from the centre of the QD outwards generates a ring structure around a crater \cite{luyken2001growth}. 

It is essential to note that this transition happens anistropically. QRs often have an elongated radial profile \cite{raz2003formation, kuroda2005optical} or an azimuthally dependent height \cite{fomin2007theory}. Whilst there has been some theoretical work which takes into account the range of asymmetries that can arise in QRs \cite{vinasco2018effects}, most models employ simple geometries, such as one- or two-dimensional rings \cite{fischer2009exciton, alexeev2012electric, govorov2002polarized, fischer2009exciton, hartmann2019uniaxial}, and do not account for deviations away from an idealised geometry. 
In this paper we include various asymmetries that are observed in QRs. However to provide a foundational understanding of the properties of QRs we initially model an idealised QR geometry, without the added complexity of defects. This simulation takes the dimensions given by Lin \textit{et al.}., that is: an inner radius $R_{\text{in}} = 15$nm, an outer radius $R_{\text{out}} = 30$nm. However, we elect to use a maximum height $h_{0} = 2$nm (instead of the estimated 1nm) to align the ground state energy with that measured in the PL spectra \cite{lin2009shape, lin2009temperature}. The radial height profile, between the inner and outer radius, of the QR is given by $h(r) = h_{0} \sin(\pi (r - R_{\text{in}}) / (R_{\text{out}} - R_{\text{in}}))$. This geometry results in a ground state exciton energy of 1255meV.

Simulations with this geometry reveal that the electron and hole wavefunctions are not delocalised across the entire QR, as is typically found in theoretical models of QRs \cite{tomic2006parallel, teodoro2010aharonov, hartmann2019uniaxial}. Rather the electron and the hole are concentrated in one particular region of the QR. The possibility of the electron and hole localising in one part of the QR has been previously investigated \cite{govorov2002polarized, kim2017quasi}. The mean circumference of the QR is almost an order of magnitude larger than the Bohr radius of InAs (which is 35nm \cite{madelung2004semiconductors}). This inhibits the electron from entirely delocalising around the QR. 

The localisation of the electron and hole explain an important feature of the PL spectrum of QRs. That is, the linewidth of the ground state emission is relatively narrow \cite{lin2009temperature, lin2009shape}. This is not the PL spectrum that is predicted based on simple ring models of QRs. These models produces states that commute with the angular momentum operator and, hence, the wavefunctions of the electron and hole are separable into vertical, radial and azimuthal parts of the form: $\psi_\ell(r, \theta, z) = R(r) Z(z) e^{i \ell \theta}$, where $\ell$ corresponds to the angular momentum of the state. As we do not expect the emitted light to carry any angular momentum this enforces the selection rule for QRs that only excitons of zero total angular momenumtum can radiatively recombine ($\ell_e + \ell_h = 0$). This implies that many excitons should contribute to the PL spectrum e.g. ($\psi_e, \psi_h$) = ($\psi_0$, $\psi_0$), ($\psi_1$, $\psi_{-1}$), ($\psi_{-1}$, $\psi_1$), ($\psi_2$, $\psi_{-2}$), etc. As each of these states are typically a few meVs apart this implies that the PL spectrum should exhibit a very broad peak emerging from the many states near the ground state that can radiatively recombine. As our simulations reveal, QRs do not have states that commute with the angular momentum and, therefore, we expect different selection rules and a different PL spectrum. 

We now consider a QR with defect that enlarges a portion of it. The QRs that Lin fabricates make no mention of elongation or specific azimuthally dependent height function. What is apparent from the AFM image is a significant number of imperfections in most of the QRs \cite{lin2009temperature, lin2009shape}. We systematically explore the effect that different sorts of defects have on the properties of the QRs.

We model defects by increasing the height and width of the QR for a portion of the ring. One half of the QR is extended outwards by following the radius of an ellipse with a maximum distance $w$ beyond the circular radius $R_{\text{out}}$. The expression defining the outer radius for QRs with this sort of defect is
\begin{equation}
	R_{\text{out}}^{(\text{def})}(\theta) = 
	\begin{cases}
		&\frac{R_{\text{out}} + w}{\sqrt{1 + \frac{w}{R_{\text{out}}^2}(2R_{\text{out}} + w)\cos^2(\theta)  }}, \; \theta \leq \pi \\
		&R_{\text{out}}, \; \theta > \pi \; .
	\end{cases}
\end{equation}

The height profile follows the same formula as previously except it is mulitplied by azimuthally dependent factor to account for the uneveness in its height
\begin{equation}
	h^{(\text{def})}(r, \theta) = 
	\begin{cases}
		&h(r)(1 + A \sin^2(\pi \theta / \theta_0)  ), \; \theta \leq \theta_0 \\
		&h(r), \; \theta > \theta_0
	\end{cases}
\end{equation}

\begin{figure}
	\includegraphics[width=\linewidth]{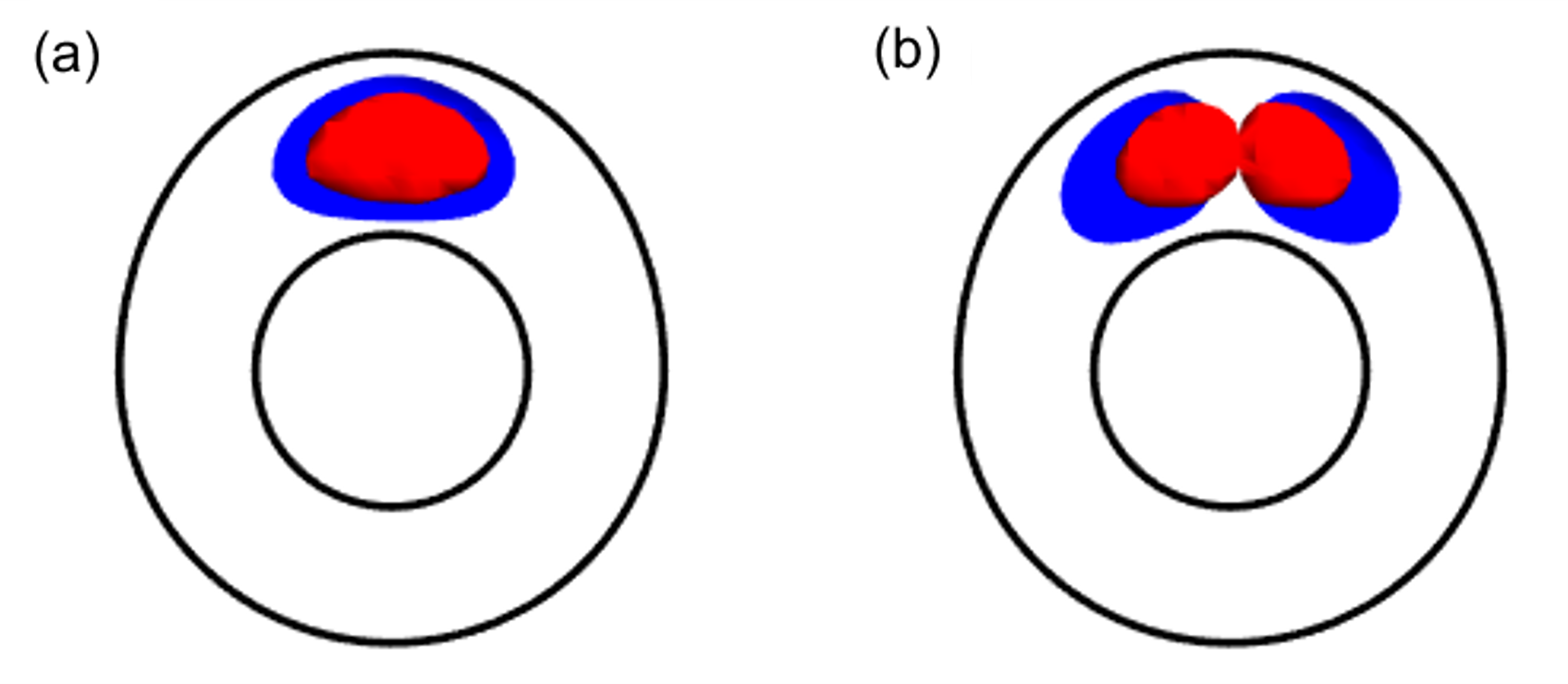}
	\caption{The electron (blue) and hole (red) wavefunctions for (a) the ground state exciton and (b) the first excited state exciton. The two excitons have 1247meV and 1293meV, closely matching the experimentally observed bright state energies.}
	\label{fig:wfs}
\end{figure}

\begin{figure}
	\includegraphics[width=\linewidth]{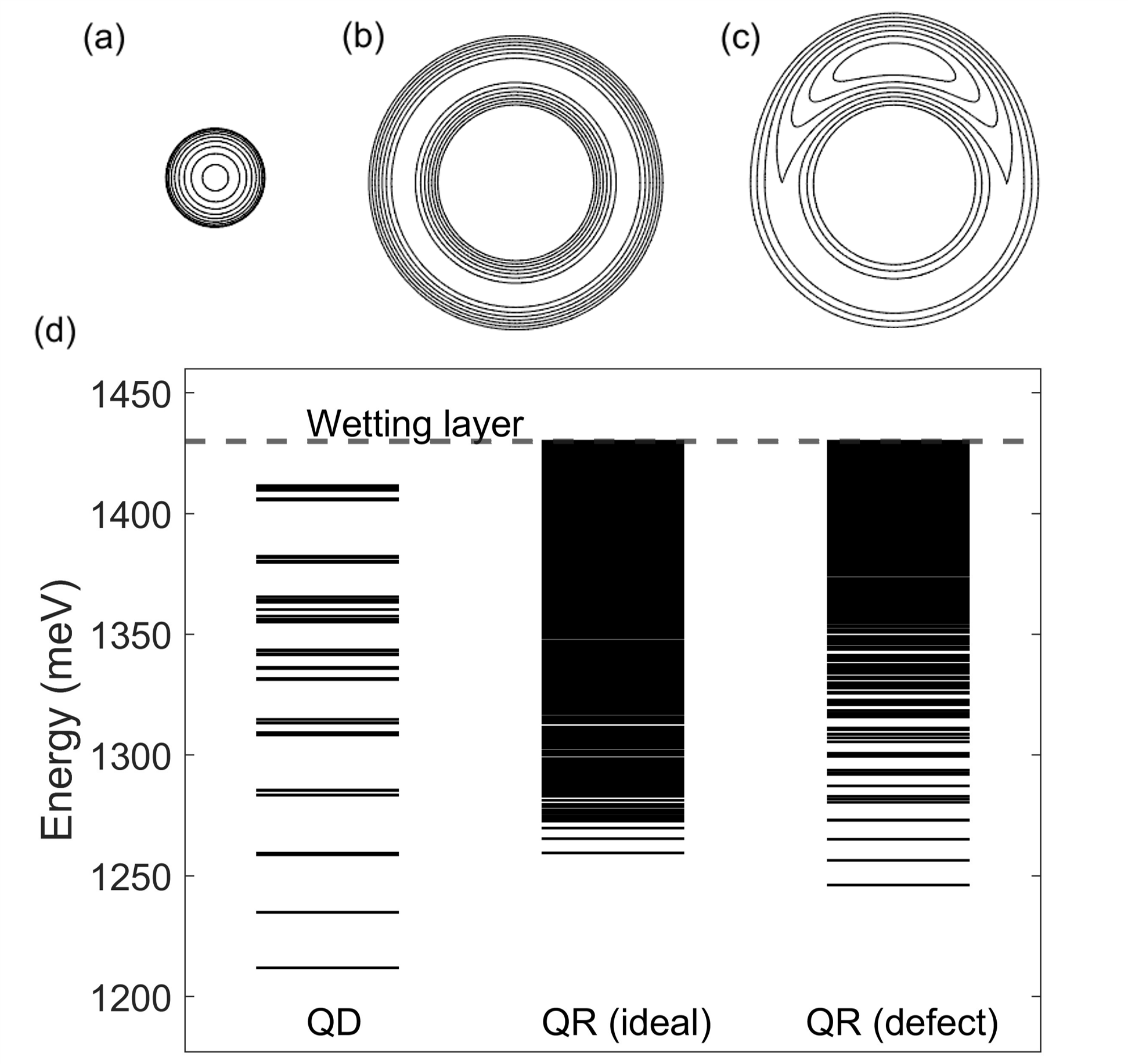}
	\caption{The geometry of (a) the QD, (b) the idealised QR and (c) the QR with a defect. The height of each structure is presented as a contour plot where each contour is vertically separated by 0.2nm. The QR with the defect has the parameters: $A = 1$, $\theta_0 = \pi$ and $w = 5$nm. (d) The energy spectra of each structure where each line represents an excitonic state at a particular energy.}
	\label{fig:contour}
\end{figure}

With this formulation of the defect, there exists a range of parameters that result in very good agreement with the observed PL spectrum of the QR. For instance, using the parameters $A = 1$, $\theta_0 = \pi$ and $w = 5$nm results in a ground state exciton of 1247meV and a bright exciton at approximately 1293meV. These two energies closely match to energies of the two bright states found in the PL spectrum. The wavefunctions of these two excitons are demonstrated in Fig. \ref{fig:wfs}. The geometry of this structure, as well as the QD and idealised QR, are presented in Fig. \ref{fig:contour}. 

We also used the FEM to calculate the piezoelectric effect. We found that for the dimensions of QRs investigated in this paper it had little to no effect. This is consistent with other work that found QRs with large ratios between their outer radius and height would not be significantly influenced by the piezoelectric effect \cite{barker2004electron}.

\section{The temperature-lifetime model}
The relationship between the exciton lifetime of a quantum structure and the temperature depends on its spectrum of states, whether those states are bright or dark and thermal escape mechanisms. Despite this relationship being affected by many parameters and despite there being many different types of QDs, most QDs exhibit approximately the same temperature-lifetime curve \cite{yang1997effect, heitz1999temperature, sauvage2002long, de2006size,harbord2009radiative, oron2009universal}. At very low temperatures (less than 50K) it is observed that the lifetime of the system matches the lifetime of the ground state exciton, typically around 1ns. This occurs as the system is primarily in the ground state and thermal processes are largely suppressed. Increasing the temperature (typically to around 100K) results in slight increase in the lifetime of the QD. This is because the dark states, that are typically tens of meV above the ground state, begin to become occupied and delay radiative recombination. This process is ultimately suppressed at even higher temperatures as thermal escape leading to non-radiative recombination comes to dominate the QD. The increasing contribution from non-radiative recombination is reflected in the reduced PL observed at higher temperatures. Thermal escape reduces the lifetime of the QD until, at approximately room temperature, the lifetime of the QD is effectively unchanged by further increases to temperature.

The relationship for QRs is quite different. At low temperatures, the QR's lifetime approximately matches the lifetime of the QD. However, from approximately 150K, instead of the lifetime decreasing there is marked increase in liftetime to about 10.5ns \cite{lin2009shape, lin2009temperature}.

This phenomenon is attributed to the different spectra of states in QRs and QDs. The QR has many more states close to the ground state which are assumed to be dark states. As the temperature rises, these dark states become thermally populated thereby increasing the lifetime of the QR. This theory is supported by a two-level rate equation model \cite{lin2009shape, lin2009temperature}.

We construct a more sophisticated model that takes into account the full spectrum of bright and dark excitonic states in each quantum structure. Our model, like other similar models, assumes that the total decay time can be attributed to the radiative and non-radiative lifetimes \cite{gurioli1992thermal, harbord2009radiative, man2015discrete}
\begin{equation}
	\frac{1}{\tau} = \frac{1}{\tau_r}  + \frac{1}{\tau_{nr}}   \; ,
\end{equation}
where the radiative and non-radiative lifetimes of the structure are $\tau_r$ and $\tau_{\text{nr}}$, respectively.

The radiative lifetime of a quantum structure can be calculated by considering the probability each excitonic state is occupied and the probability of that state radiatively recombining. By assuming that the probability of a particular excitonic state being occupied is given by the Boltzmann distribution, the following expression can be given for the radiative lifetime \cite{takagahara1993nonlocal, gotoh1997radiative, harbord2009radiative}.
\begin{equation} \label{eq:rad}
	\frac{1}{\tau_r} =  \frac{\sum_j^N \Gamma_j e^{-E_j / k_B T} }{Z} \; ,
\end{equation}
where $\Gamma_j$ is the radiative recombination rate of the $j^{\text{th}}$ excitonic state, $Z$ is the partition function and $N$ is the number of excitonic states the quantum structure supports.

Our formulation of the radiative lifetime follows the work of Harbord \textit{et al.} \cite{harbord2009radiative}, however, we extend it by accounting for electron and hole states separately. The probability that an exciton composed of an electron in state $i$ and hole in the state $j$ (which we label as $e_ih_j$) will recombine is given by Fermi's golden rule \cite{harrison2016quantum}
\begin{equation}
	\Gamma_{ij} = \frac{2 \pi}{\hbar} |\langle i | H | j \rangle|^2 \; .
\end{equation}

The matrix element for interband transitions in semiconductors is given by the following expression
\begin{equation}
	\langle i | H | j \rangle = p_{cv} \langle i | j \rangle
\end{equation}
where $ p_{cv}$ is the interband momentum matrix element. 

We can relate the radiative rate of the exciton $e_ih_j$ to the radiative rate of the ground state exciton $e_1h_1$:
\begin{equation}
	\Gamma_{ij} = \Gamma_{11}\frac{|\langle i | j \rangle|^2 }{|\langle 1 |1 \rangle|^2}   =  \frac{1}{\tau_0}\frac{|\langle i | j \rangle|^2}{ |\langle 1 |1 \rangle|^2}  \; ,
\end{equation}
where $\tau_0$ is the lifetime of the ground state exciton.

We may re-define the radiative lifetime of the system, previously given by (\ref{eq:rad}), in terms of the electron and hole states.
\begin{equation} 
	\frac{1}{\tau_r} =   \frac{1}{Z_e Z_h \tau_0 |\langle 1 |1 \rangle|^2}  \sum_{i}^{N_e} \sum_{j}^{ N_h} |\langle i | j \rangle|^2 \; e^{-(E^{(e)}_i + E^{(h)}_j) / k_B T}   
\end{equation}
where the partition function for the electron and hole are defined as
\begin{align}
	Z_e &=   \sum_{i}^{N_e} e^{-E^{(e)}_i / k_B T}  \\
	Z_h &=   \sum_{j}^{N_j} e^{-E^{(h)}_i / k_B T} \; .
\end{align}

To simplify the calculation we set the zero energy for the electron and hole to correspond to the ground state of each charge carrier. In the case where the only bright exciton is the ground state exciton, the radiative lifetime reduces to a particularly simple expression
\begin{equation} 
	\tau_r =  Z_e Z_h \tau_0  \; .
\end{equation}

We calculate the non-radiative lifetime of quantum structures in a similar manner. Our derivation of the non-radiative lifetime follows that of Gurioli \textit{et al.} \cite{gurioli1992thermal}. However, we adapt their expression to account for the separate electron and hole states. We assume non-radiative processes predominantly occur when an exciton thermally escapes into the wetting layer. The likelihood that an electron or hole can make this jump depends on the difference between its energy and the energy of the wetting layer. This difference is the activation energy. However, it has been demonstrated that for different quantum structures the activation energy may be modified according to whether thermal escape involves excitons, correlated electron-hole pairs or single charge carriers \cite{bacher1991influence, yang1997effect, khatsevich2005cathodoluminescence, schulz2009optical, gelinas2010carrier, munoz2012size, jahan2013temperature}. We use the factor $\nu$ to account for the different types of thermal emission which are possible. The total non-radiative rate is determined by the thermal escape rate of each exciton state and the probability of each state being occupied (which we assume is given by the Boltzmann distribution)
\begin{align}
	\begin{split}
		\frac{1}{\tau_{nr}} &=   \Gamma_{0} \sum_{i}^{N_e} \sum_{j}^{ N_h}   e^{-(\nu E^{(e)}_{WL} - E^{(e)}_{i} ) / k_B T} \; e^{-(\nu E^{(h)}_{WL} - E^{(h)}_{j} ) / k_B T}  \\
		&\times     \frac{e^{-E^{(e)}_i / k_B T} }{ Z_e  }     \frac{ e^{-E^{(h)}_j / k_B T}}{  Z_h }       \\
		&=  \frac{\Gamma_{0}}{ Z_e Z_h }   \sum_{i}^{N_e} \sum_{j}^{ N_h} e^{-\nu (E^{(e)}_{WL} + E^{(h)}_{WL} ) / k_B T}     \\
		&=  \frac{\Gamma_{0} N_e  N_h }{ Z_e Z_h }  e^{-\nu E_{WL} / k_B T} 
	\end{split}
\end{align}
where the wetting layer energies for the electron and hole are $E_{WL}^{(e)}$ and $E_{WL}^{(h)}$, respectively. The total exciton wetting layer energy is $E_{WL} = E_{WL}^{(e)} + E_{WL}^{(h)}$. The scattering rate from confined states to the wetting layer is $\Gamma_{0}$.

Using the formalism presented here and the FEM results of the previous section we are able to calculate the temperature-lifetime relationship of each structure and its dominant thermal escape mechanism.

\section{The interplay of geometry and lifetime}
The model described in the previous section requires knowledge of the energy levels of the quantum structure being investigated. By combing this model with our FEM results we are able to determine the temperature-lifetime relationship of each structure. We find that this relationship is sensitive to the geometry. This is because each structure's spectrum depends on its specific shape. The spectra of the QD, idealised QR and QR with a defect are displayed in Fig. \ref{fig:contour}.

The temperature-lifetime relationship we calculate for QDs matches the experimentally observed relationship when $\nu = 0.5$. This corresponds to thermal escape primarily by correlated electron-hole pairs, which is consistent with other experiments on QDs \cite{yang1997effect, schulz2009optical, gelinas2010carrier, jahan2013temperature}.

Our results also demonstrates that the QR with the defect, described in the previous section, has a temperature-lifetime relationship that approximately matches the experimentally observed relationship for the QR \cite{lin2009temperature, lin2009shape}. This is in stark contrast to the relationship exhibited by the idealised QR which exhibits a significant rise in its lifetime at a temperature much lower than is experimentally observed. It is worth clarifying why each QR produces a distinct temperature-lifetime relationship.

\begin{figure}
	\includegraphics[width=\linewidth]{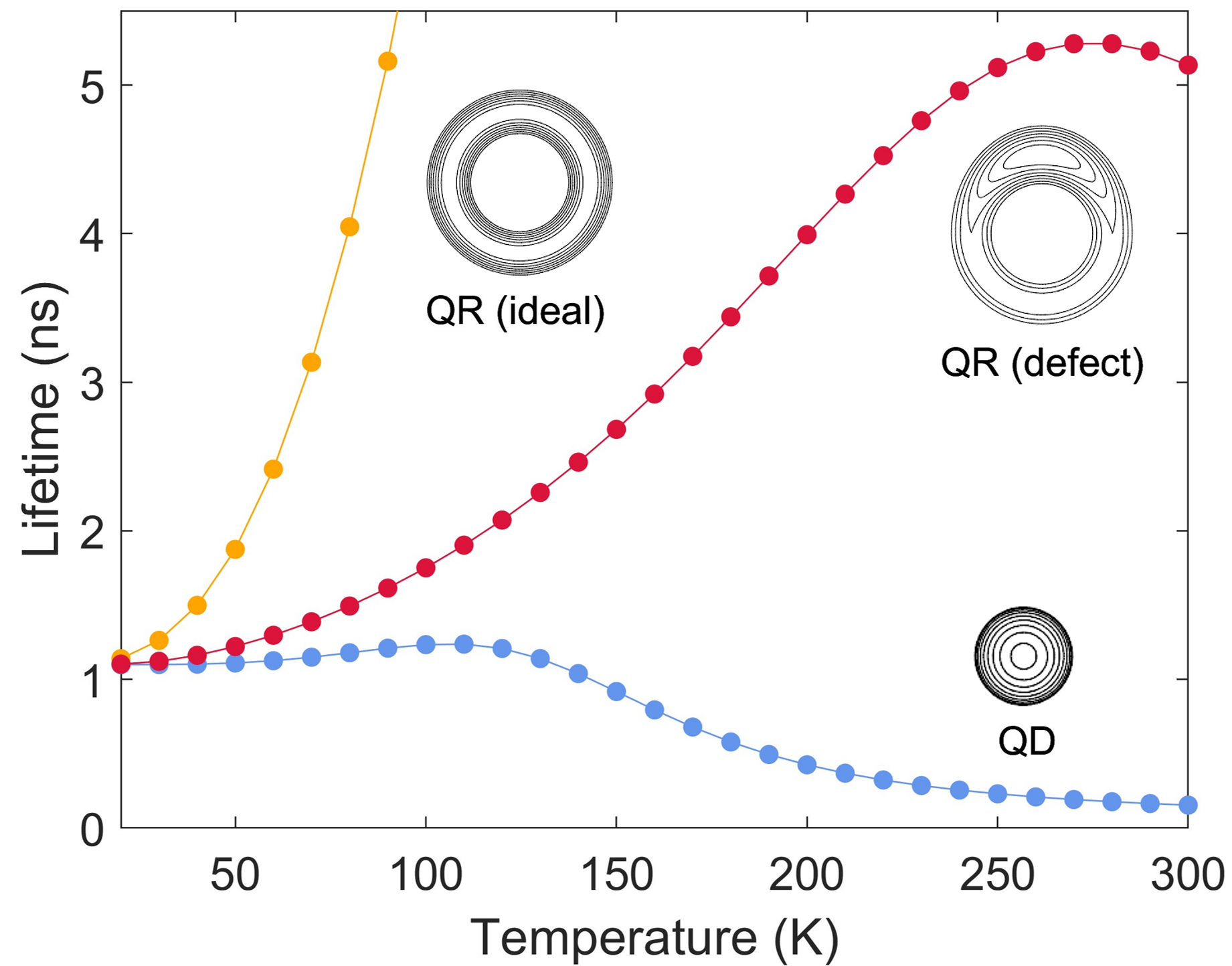}
	\caption{The lifetimes of the QD, the QR with a defect and an idealised QR. The emission parameter is $\nu = 0.5$ for the QD and $\nu = 1$ for the QRs. The thermal emission rate is $\Gamma_0 = 10$ for the QD and $\Gamma_0 = 2$ for the QRs.}
	\label{fig:lifetimes}
\end{figure}

The marked increase in lifetime at much low temperatures for the idealised QR can be explained by the proximity of dark states to the ground state as was suggested by Lin \textit{et al.} \cite{lin2009shape, lin2009temperature}. The spectrum of the idealised QR is illustrated in Fig. \ref{fig:contour}, revealing the many states close to its ground state. Minimal increases in temperature causes these states to become occupied and, since they are dark states, the occupation of these states increases the lifetime of the system.

The QR with the defect has a significant spacing between its ground state and the next excited state. Its energy spectrum is effectively that of a QD at low energies and that of the idealised QR at higher energies. The consquence of this is that at low temperatures the ground state is primarily occupied, just like in the QD. This results in a large range of temperatures where the QR is most likely in the ground state and, hence, there is only a marginal change to the lifetime in this range. However, once the temperature is reached where the many dark states can be accessed, then there is a sudden increase in lifetime. The defect in the QR provides significant energy spacing around the ground state and the ring geometry provides a close packed set of states at higher energies. Only a spectrum of states with both these features can produce the temperature-lifetime relationship observed for QRs \cite{lin2009temperature, lin2009shape}.

\begin{figure}
	\includegraphics[width=\linewidth]{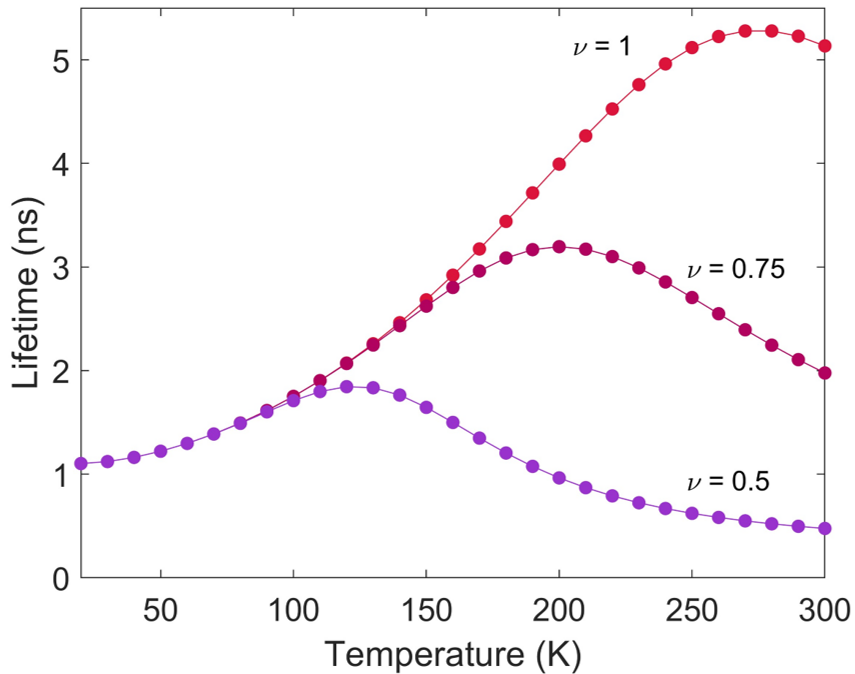}
	\caption{The lifetimes of the QR with a defect for different $\nu$ values. Only when $\nu \approx 1$ does this relationship match what is experimentally observed.}
	\label{fig:nu}
\end{figure}

We now consider the type of thermal emission present in QRs. To fit to the experimental results \cite{lin2009shape, lin2009temperature} requires $\nu = 1$, implying that the thermal emission from QRs involves the ejection of excitons. The temperature-lifetime relationship of QRs with different values of $\nu$ is depicted in Fig. \ref{fig:nu}. The difference in thermal escape mechanisms for QDs and QRs suggests that geometric effects are an important factor in the process. It should be noted that elsewhere it has been calculated that the geometry of QDs significantly effects its capture processes \cite{magnusdottir2002one, magnusdottir2003geometry}.

\begin{figure}
	\includegraphics[width=\linewidth]{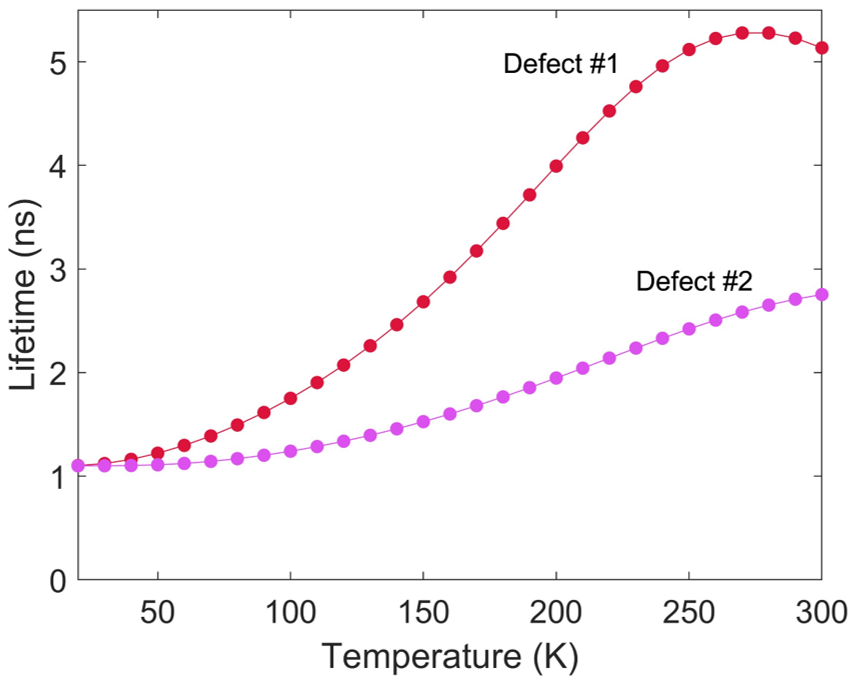}
	\caption{The lifetime-temperature relationship of two QRs with different defects. The first defect is characterised by $A = 1$, $\theta_0 = \pi$ and $w=5$; the second defect has the parameters $A = 1.25$, $\theta_0 = \pi/2$ and $w=0$. We have assumed thermal escape via excitons ($\nu = 1$) for both structures.}
	\label{fig:defects}
\end{figure}

There are some quantitative differences between the temperature-lifetime relationships we calculate and the one measured for QRs. We find that the lifetime of the QR diverges from than the lifetime of QD at a lower temperature than is experimentally observed \cite{lin2009temperature, lin2009shape}. The increased lifetime that we find for the QR arises from the lowest energy dark states being energetically too close to the ground state. However, this sort of spectra did not arise for every defect we investigated. For instance, if we consider a QR with height defect parameters of $A = 1.25$ and $\theta_0 = \pi/2$ but no change to the width ($w = 0$), we find a similar ground state exciton energy of 1245meV. Yet this QR has a temperature-lifetime relationship that approximately matches that of the QD up to 150K, as is experimentally observed for QRs \cite{lin2009temperature, lin2009shape}. The maximum lifetime of this QR only reaches 2.75ns, however, which is well short of the 10.5ns observed. The temperature-lifetime relationships for these two QRs are demonstrated in Fig. \ref{fig:defects}. 

Nearly every QR with a defect that we modelled, with a ground state energy that approximately matched the measured PL spectrum, had a temperature-lifetime relationship between the two curves in Fig. \ref{fig:defects}. To obtain high precision, quantitative agreement would require a detailed model of the materials and any defects, but qualitatively the trends observed are general.

\section{Conclusion}
In this paper we demonstrate that the distinctive temperature-lifetime relationship observed for QRs is a product of its spectrum of states and its dominant thermal escape mechanism. The spectrum of states required to produce the observed lifetime requires siginificant energy spacing between states near the ground state but at higher energies closely packed states. Rotationaly symmetric models of QRs cannot generate this set of states but QRs with defects that enlarge a portion of the ring can. 
 
In matching to the experimentally observed temperature-lifetime relationship, we find that the dominant thermal escape mechanism in QRs involves the ejection of excitons whereas the QDs we investigated thermally emit correlated electron-hole pairs. We, therefore, suggest that the geometry plays an important role in the thermal escape processes in a quantum structure.

The nature of thermal escape from nanostructures is an ongoing question, but our results suggest the geometric effects are key to understanding this behaviour. 
\bibliography{bibfile}

\end{document}